\documentclass[reprint,superscriptaddress,amsmath,amssymb,
 aps,showkeys]{revtex4-2}
\usepackage{graphicx}
\usepackage{dcolumn}
\usepackage{bm}
\usepackage{multirow}
\usepackage{booktabs}
\usepackage{float}
\usepackage{placeins}
\usepackage{xcolor}
\usepackage[colorlinks=true,linkcolor=blue,citecolor=blue,urlcolor=blue]{hyperref}
\usepackage{miller}

\begin{document}

\preprint{APS/123-QED}

\title{Thermally activated nature of synchro-Shockley dislocations in Laves phases} 

\author{Zhuocheng Xie}
\email[]{xie@imm.rwth-aachen.de}
\affiliation{Institute of Physical Metallurgy and Materials Physics, RWTH Aachen University, 52056 Aachen, Germany}

\author{Dimitri Chauraud}
\affiliation{Max-Planck-Institut für Eisenforschung GmbH, Max-Planck-Str. 1, 40237 Düsseldorf, Germany}
    
\author{Achraf Atila}
\affiliation{Department of Materials Science and Engineering, Institute I: General Materials Properties, Friedrich-Alexander-Universität Erlangen-Nürnberg, 91058 Erlangen, Germany}
\affiliation{Department of Materials Science and Engineering, Saarland University, 66123 Saarbrücken, Germany}
     
\author{Erik Bitzek}
\affiliation{Max-Planck-Institut für Eisenforschung GmbH, Max-Planck-Str. 1, 40237 Düsseldorf, Germany}

\author{Sandra Korte-Kerzel}
\affiliation{Institute of Physical Metallurgy and Materials Physics, RWTH Aachen University, 52056 Aachen, Germany}

\author{Julien Guénolé}
\email[]{julien.guenole@univ-lorraine.fr}
\affiliation{Université de Lorraine, CNRS, Arts et Métiers ParisTech, LEM3, 57070 Metz, France}
\affiliation{Labex Damas, Université de Lorraine, 57070 Metz, France}

\begin{abstract}
Synchro-Shockley dislocations, as zonal dislocation, are the major carrier of plasticity in Laves phases at high temperatures. The motion of synchro-Shockley dislocations is composed of localized transition events, such as kink-pair nucleation and propagation,  which possess small activation volumes, presumably leading to sensitive temperature and strain rate dependence on the Peierls stress. However, the thermally activated nature of synchro-Shockley dislocation motion is not fully understood so far. In this study, the transition mechanisms of the motion of synchro-Shockley dislocations at different shear and normal strain levels are studied. The transition processes of dislocation motion can be divided into shear-sensitive and -insensitive events. The external shear strain lowers the energy barriers of shear-sensitive events. Thermal assistance is indispensable in activating shear-insensitive events, implying that the motion of synchro-Shockley dislocations is prohibited at low temperatures.
\end{abstract}


\maketitle

Laves phases, as the most common intermetallic phases, exist in many alloys and have a great impact on their mechanical properties due to their high strength and good creep resistance compared to the matrix phases \cite{sinha1972topologically,paufler2011early,stein2021laves}. Although Laves phase alloys often exhibit excellent mechanical properties at high temperatures, their notorious brittleness at room temperature limits their structural applications \cite{stein2021laves,livingston1992laves,pollock2010weight}. Understanding the underlying deformation mechanisms of Laves phases is thus crucial for tailoring the material properties of the composites.

Synchro-shear is the dominant mechanism for dislocation-mediated plasticity on the basal or $\{ 1 1 1 \}$ plane in Laves phases,
which was confirmed by experimental observations of synchro-shear-induced stacking faults and synchro-Shockley dislocation cores in C14 HfCr\textsubscript{2} \cite{chisholm2005dislocations}. Moreover, as demonstrated by $ab$ $initio$ calculations \cite{vedmedenko2008first, luo2023Tailoring} and atomistic simulations \cite{guenole2019basal}, synchro-shear slip is the energetically favorable mechanism compared to another competitive crystallographic slip between kagomé and triple layers for basal or $\{ 1 1 1 \}$ planes in Laves phases.

As a typical zonal dislocation \cite{kronberg1957plastic, anderson2017theory}, the synchro-Shockley dislocation consists of two coupled Shockley partial dislocations on the adjacent planes of a triple-layer \cite{hazzledine1992synchroshear}. The motion of the synchro-Shockley dislocation requires the cooperative motion of these two coupled Shockley partial dislocations.
Recently, the authors reported the mechanisms of motion of synchro-Shockley dislocations in C14 CaMg\textsubscript{2} and C15 CaAl\textsubscript{2} using atomistic simulations \cite{xie2022unveiling}.  Two 30$^{\circ}$ synchro-Shockley dislocations, partial~I and II, with symmetric Burgers vectors along dislocation lines but different core structures and energies, were identified. The nudged-elastic band (NEB) calculations illustrated that the motion of partial~I and II dislocations are composed of nucleation and propagation of kink-pairs and short-range atomic shuffling processes, respectively.

Laves phases tend to be ductile only at elevated temperatures, where the thermal fluctuation is significant for activating synchro-Shockley dislocation motion \cite{livingston1992laves,yoshida2002tem,kazantzis2007mechanical}.
In strain-rate jump compression tests on bulk C15 NbCr\textsubscript{2} at high temperatures, when the strain rate changes, variations in the mobile dislocation density in addition to the delayed responses of dislocation velocity were reported \cite{kazantzis2008self}. These phenomena were attributed to the self-pinning nature of synchro-Shockley dislocations, namely synchro-Shockley dislocations can be temporarily pinned if one of the two coupled Shockley components is hindered due to unsuccessful thermal activation \cite{kazantzis2008self}.

\begin{figure*}[htbp!]
\centering
\includegraphics[width=0.85\textwidth]{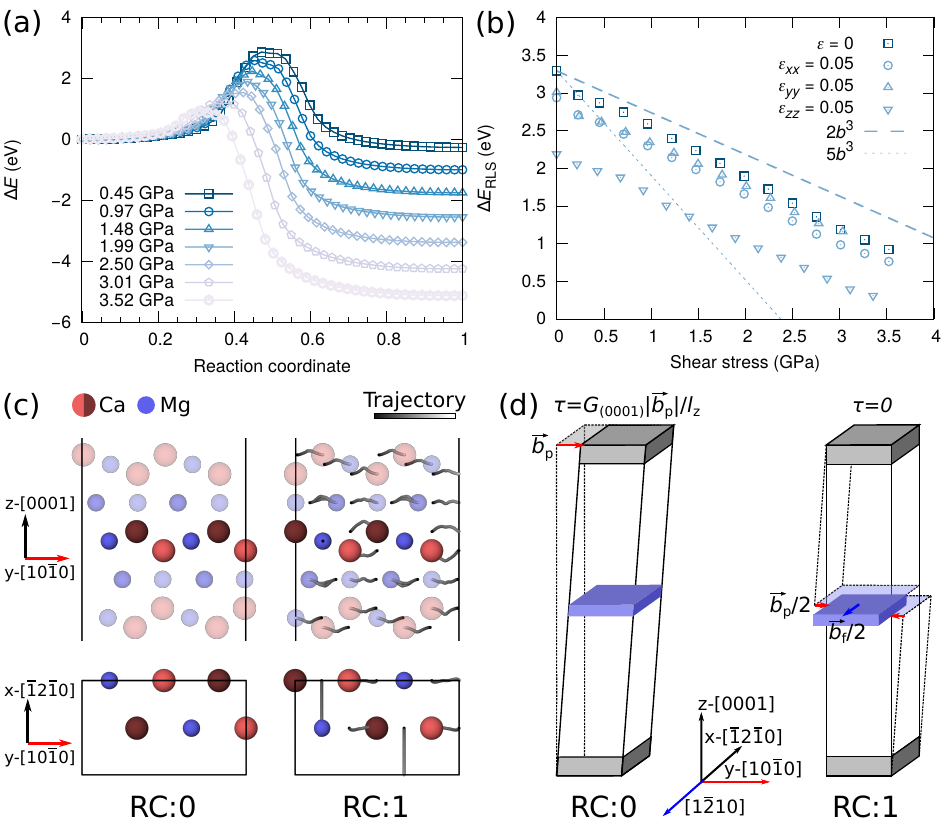}
\caption{Synchro-shear slip in C14 CaMg\textsubscript{2} at different applied stress states. (a) Shear stress-dependent minimum energy paths for leading partial of synchro-shear slip under zero normal stress. (b) Shear stress-dependent activation energy of rate-limiting step (RLS) of synchro-shear slip under zero and 0.05 uniaxial strain $\epsilon_{ii}$. (c) Snapshots of the synchro-shear slip mechanisms at reaction coordinate (RC):0 ($\tau =$ 0.45 GPa, $\epsilon_{ii} =$ 0) and RC:1 ($\tau =$ 0 GPa, $\epsilon_{ii} =$ 0). Only atoms in the triple-layer where the synchro-shear slip takes place (highlighted) are shown in the view along $[0001]$ (bottom). Large (red) and small (blue) spheres are Ca and Mg atoms, respectively. Dark and light red atoms indicate Ca atoms in different atomic layers of the triple-layer. Grey-scale lines indicate the trajectories of atoms colored according to the reaction coordinates. (d) Schematics of deformation of Laves crystal introduced by synchro-shear slip. Left: initial configuration under pre-shear strain $\epsilon_{yz}^{pre} =$ $b_\textsubscript{p}/l_{z}$, where $b_\textsubscript{p}$ is the vector of partial slip and $l_{z}$ is the dimension of the setup in $z$. The applied shear stress $\tau =$ $G_\textsubscript{(0001)}b_\textsubscript{p}/l_{z}$, where $G_\textsubscript{(0001)}$ is the shear modulus of the basal (0001) plane. Right: final configuration after the synchro-shear slip and release of stored elastic energy of pre-shear strain. The blue block indicates the Mg atomic layer in the triple-layer with the transverse displacement of $b_\textsubscript{f}/2$ after the synchro-shear slip, where $b_\textsubscript{f}$ is the vector of full slip. The transverse displacement vectors and corresponding crystallographic orientations are indicated in the coordinate system. The semi-fixed boundary conditions where atoms are frozen in shear and non-periodic directions are marked in grey.}
\label{fig1}
\end{figure*}

So far, the thermally activated nature of synchro-Shockley dislocations is not fully understood. In this study, the transition mechanisms of the rate-limiting steps (RLS) of the motion of synchro-Shockley dislocations at different shear and normal strain levels are studied using the NEB method \cite{henkelman2000climbing,henkelman2000improved}. By definition, NEB calculations are performed at 0 K, and transition state theory \cite{vineyard1957frequency} can be used to predict the behavior at elevated temperatures. The stress-dependent activation energy and volume of dislocation motion were investigated and correlated to thermal activation. 

Atomistic simulations were performed using LAMMPS \cite{LAMMPS}. 
The interatomic interactions were modeled by the modified embedded atom method potential by Kim et al. \cite{kim2015modified} for Mg-Ca. The synchro-Shockley partial dislocations (partial~I: 1/3$[ 1 0 \bar{1} 0 ]$; partial~II: 1/3$[ 0 1 \bar{1} 0 ]$) in C14 CaMg\textsubscript{2} were constructed following the method detailed in \cite{xie2022unveiling,hirel2015atomsk,vaid2022pinning}, and then relaxed using the conjugate gradient with box relaxation and the FIRE \cite{bitzek2006structural,guenole2020assessment} algorithm. OVITO \cite{stukowski2009visualization} was used for visualization and analysis. The applied shear stress $\tau$ value was derived from the shear modulus on the basal plane given by the elastic constant $C_\textsubscript{44}$ of the interatomic potential ($G_\textsubscript{(0001)} =$ 25.6 GPa) and the applied shear strain \cite{xie2022unveiling}. For more details of methods, see Figure S1 and \cite{guenole2019basal,xie2022unveiling}.

\begin{figure*}[htbp!]
\centering
\includegraphics[width=0.85\textwidth]{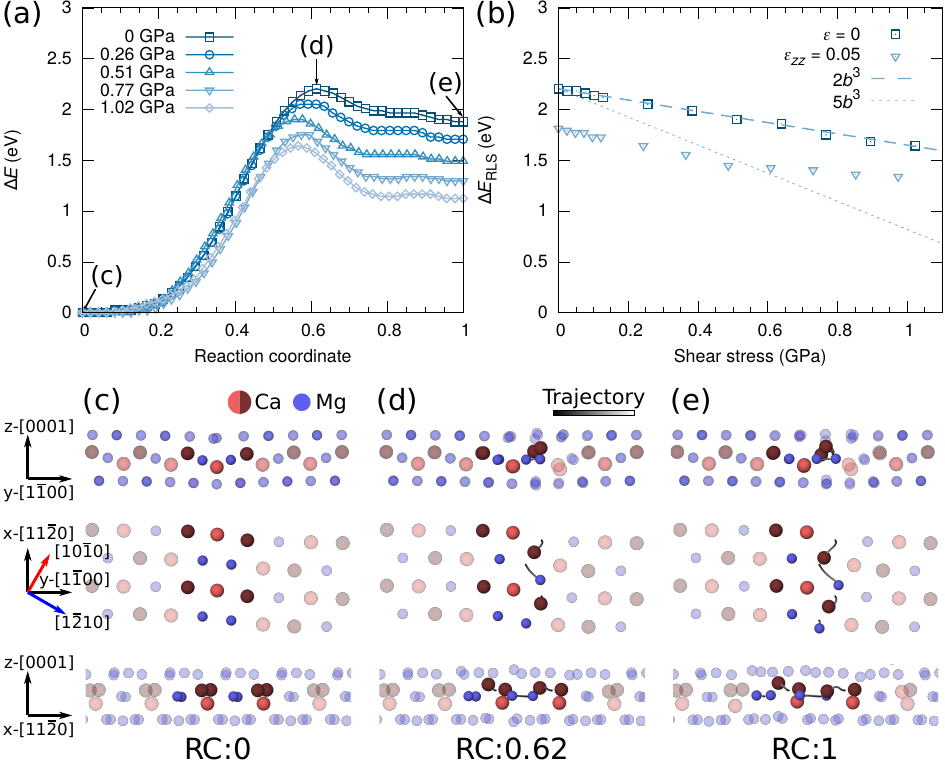}
\caption{Kink-pair nucleation of partial~I dislocation in C14 CaMg\textsubscript{2} at different applied shear and normal stress states. (a) Shear stress-dependent minimum energy paths for kink-pair nucleation under zero normal stress. (b) Shear stress-dependent activation energy of RLS (kink-pair nucleation) of motion of 30$^{\circ}$ synchro-Shockley partial~I dislocation under zero and 0.05 uniaxial strain $\epsilon_{zz}$. (c-e) Snapshots of the mechanisms of kink-pair nucleation at zero applied shear and normal stress. The same visualization and color coding schemes as Figure \ref{fig1}(c) are used here. Only the trajectories of highlighted atoms are shown here.}
\label{fig2}
\end{figure*}

The initial configuration of the NEB calculation on the synchro-shear slip is pre-strained by a transverse displacement corresponding to the partial slip vector $b_\textsubscript{p}$, as illustrated in Figure \ref{fig1}(d). The pre-strain acts as a driving force for synchro-shear slip, and its contribution to the stored elastic energy is fully released in the final configuration after the slip event. 
To investigate the stress-dependent synchro-shear slip, the strain-controlled NEB calculations were performed by superimposing additional shear strain in $b_\textsubscript{p}$ direction on both initial and final configurations.
A synchro-shear full slip involves a leading component with the creation of a stacking fault and a trailing component with the elimination of the stacking fault. Here, we focus on the leading partial slip as the trailing counterpart exhibits a mirror symmetric minimum energy path (MEP) when the external stress is zero, see Figure S2. 

The activation energy of synchro-shear slip decreases with increasing applied shear stress as shown in Figure \ref{fig1}(a). Similar mechanisms of synchro-shear slip are observed at different stress levels, namely, the upper part of the crystal (including the Ca atomic layer colored in dark red) shifts by a displacement gradient from zero to $\frac{a_{0}}{3}[ 1 0 \bar{1} 0 ]$ relative to the lower part of the crystal (including the Ca atomic layer colored in light red) and the Mg atomic layer synchronously displaces by $\frac{a_{0}}{3}[ 1 \bar{2} 1 0 ]$ (Figure \ref{fig1}(c)). Therefore, the synchro-shear slip can be decomposed into a localized atomic shuffling event (including only small atoms in the triple-layer, blue block as indicated in Figure \ref{fig1}(d)) and a global shear straining event (excluding small atoms in the triple-layer). 
The applied shear stress $\tau$ lowers the energy barrier of the global shear deformation, but not the localized atomic shuffling in the direction perpendicular to the transverse displacement introduced by the shear deformation.

\begin{figure*}[htbp!]
\centering
\includegraphics[width=0.85\textwidth]{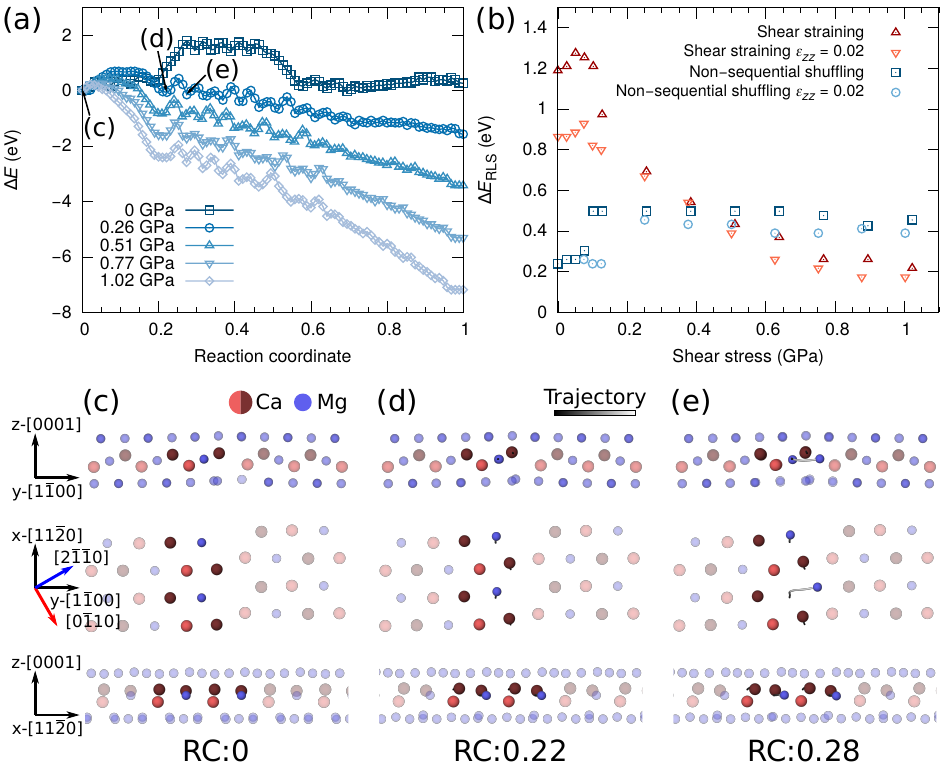}
\caption{Rate-limiting steps of motion of partial~II dislocation in C14 CaMg\textsubscript{2} at different applied shear and normal stress states. (a) Shear stress-dependent minimum energy paths for the motion of partial~II. (b) Shear stress-dependent activation energies of RLS (shear straining event when $\tau <$ 0.4 GPa and non-sequential shuffling event after shear straining stage when $\tau >$ 0.4 GPa) of motion of 30$^{\circ}$ synchro-Shockley partial~II dislocation under zero and 0.02 uniaxial strain $\epsilon_{zz}$. (c-e) Snapshots of the mechanisms of (d) shear straining and (e) non-sequential shuffling under 0.26 GPa applied shear stress and zero normal stress. The same visualization and color coding schemes as Figure \ref{fig1}(c) are used here. Only the trajectories of highlighted atoms are shown here.}
\label{fig3}
\end{figure*}

The energy barrier of the synchro-shear slip is composed of the activation energies of shear-sensitive and -insensitive events. The activation volume of the shear-sensitive component of the RLS ($\Omega_{RLS}$) is calculated by taking the first derivative of the activation energy ($E_{RLS}$) w.r.t. resolved shear stress ($\tau$),
\begin{equation}
\label{equ1}
\Omega_{RLS}=-\frac{\partial \Delta E_{RLS}(\tau)}{\partial \tau}
 ~.
\end{equation}
Here we plot 2$b^{3}$ and 5$b^{3}$, where $b$ is the partial Burgers vector with a magnitude of 3.54 \AA, as constants in Figure \ref{fig1}(b) to estimate $\Omega_{RLS}$. The $\Omega_{RLS}$ value of the shear-sensitive event of the synchro-shear slip is close to 2$b^{3}$. 

The effect of normal strain on the activation energy of synchro-shear slip at different applied shear stress states is shown in Figure \ref{fig1}(b) and Figure S3. The decrease of the energy barrier is more pronounced under the uniaxial tensile strain perpendicular to the slip plane $(0001)$ ($\epsilon_{zz}$) than the other two tensile strain directions ($\epsilon_{xx}$ and $\epsilon_{yy}$), since the atomic motion of the synchro-shear slip, particularly for the large atoms in the triple-layer, involves significant out-of-plane components (Figure \ref{fig1}(c)). Eventually, the energy barrier of the RLS of the synchro-shear slip corresponding to the localized atomic shuffling event does not vanish up to an applied shear stress $\tau =$ 3.52 GPa and uniaxial tensile strain $\epsilon_{zz} =$ 10\% (Figure S3). When $\tau >$ 3.52 GPa, a slip between the triple-layer and the adjacent kagom\'e layer other than the synchro-shear slip was observed as the first event, see Figure S4. This triple-kagom\'e slip was demonstrated as a competing mechanism to synchro-shear slip in the Laves crystal building block of $\mu$ phases in experiments and $ab$-$initio$ calculations \cite{luo2023Tailoring,luo2023plasticity}. The triple-kagom\'e full slip dissociates into two mirror symmetric partial slip events in CaMg\textsubscript{2}, and a stacking fault state with a high fault energy 409~mJ/m$^{2}$ compared to the low fault energy of the synchro-shear induced stacking fault (14~mJ/m$^{2}$, Figure S5) was obtained, see Figure S6(a-b). In addition, the energy barrier of the triple and kagom\'e slip vanishes when the applied shear stress $\tau =$ 5.39~GPa (Figure S6(c-d)). 

In Laves phases, two synchro-Shockley partial dislocations (partial~I and II) exist on the basal (or \{111\}) triple-layer. 
Our previous NEB calculations suggest that the motion of 30$^{\circ}$ synchro-Shockley dislocations is a multistep reaction that resolves into multiple transition and intermediate states \cite{xie2022unveiling}. 
For partial~I dislocation, we identified kink-pair nucleation and following kink propagation by vacancy-hopping and interstitial shuffling as the mechanisms of motion. Among these activation events, the kink-pair nucleation process exhibits the highest energy barrier, therefore is the RLS of the mechanism of dislocation motion (Figure S7(a)).  

The activation energy of kink-pair nucleation is 2.20 eV under zero applied shear and normal stress (Figure \ref{fig2}(a)). The kink-pair nucleation process involves the synchronous motion of Ca and Mg atoms in the triple-layer along 1/3$[ \bar{1} 0 1 0 ]$ and 1/3$[ 1 \bar{2} 1 0 ]$, respectively, see Figure \ref{fig2}(c-e). We calculated the shear stress-dependent excess energy of kink-pair nucleation by varying the applied shear strain in the same direction as $b_\textsubscript{p}$ on the initial and final NEB configurations (Figure \ref{fig2}(a-b) and S7(a)).
The motion of the Mg atom in the triple-layer is a short-range atomic shuffling in the slip plane perpendicular to the transverse displacement of applied shear stress and does not contribute to the global strain of the lattice, therefore it is insensitive to the applied shear stress. 
The activation energy of kink-pair nucleation decreases with the increase of applied shear stress. Note that we did not observe any significant change in the kink pair distance as the applied shear stress was increased (Figure S8). This is due to the fact that the energetic contribution of the formation of point defects is much higher than the elastic interaction energy between kinks as well as the line tension. In addition, the uniaxial tensile strain $\epsilon_{zz}$ significantly lowers the energy barrier (Figure \ref{fig2}(b) and S7(b)). However, the energy barrier of kink-pair nucleation would not vanish under the shear and uniaxial loading due to the shear-insensitive component. When $\tau >$ 1.02 GPa, a plastic event between triple and kagom\'e layers was activated other than the motion of synchro-Shockley partial dislocations. Two constants of activation volumes, 2$b^{3}$ and 5$b^{3}$ according to Equation \ref{equ1}, were plotted in Figure \ref{fig2}(b) and the estimation of $\Omega_{RLS}$ is close to 2$b^{3}$, a similar value as obtained in the simulation setup for synchro-shear slip.

For partial~II, multiple transition events including non-sequential atomic shuffling, shear straining and short-range rearrangement were identified as the mechanisms of motion \cite{xie2022unveiling}. Under zero applied shear stress, the shear straining event has the highest activation energy compared with  the other two events (Figure \ref{fig3}(a)). In addition, the activation energy of the shear straining process is proportional to the dislocation length (0.018 eV/\AA), therefore, it is neither a localized nor a thermally activated event \cite{xie2022unveiling}. 
The stress-dependent NEB calculations of partial~II were performed by applying shear strain parallel to the Burgers vector of partial~II along $[ 0 \bar{1} 1 0]$. With increasing applied shear stress from 0 to 0.1 GPa, the sequence of transition events on the MEP rearranges (Figure \ref{fig3}(a)), therefore abrupt changes of activation energies of transition events are observed (Figure \ref{fig3}(b)). When $\tau >$ 0.1 GPa, shear straining becomes the first event, followed by non-sequential atomic shuffling and short-range rearrangement. The activation energy of the shear straining event decreases rapidly with increasing applied shear stress. When $\tau >$ 0.4 GPa, the non-sequential shuffling event becomes the RLS with the highest activation energy on the MEPs, see Figure \ref{fig3}(b). In addition, the activation energy of the non-sequential shuffling process remains almost constant (around 0.5 eV) with increasing applied shear stress. Similar to the localized atomic shuffling event of kink-pair nucleation of partial~I, the non-sequential shuffling accomplished by the motion of the Mg atom in the triple-layer does not contribute to the global shear deformation, therefore it is a shear-insensitive event. The uniaxial tensile strain $\epsilon_{zz}$ lowers the energy barriers of both shear-sensitive and insensitive events (Figure \ref{fig3}(b) and S7(c)).

For synchro-Shockley dislocations, shear-sensitive and insensitive events co-exist on the MEPs of dislocation motion. As the resolved shear stress can only assist to overcome the energy barriers of shear-sensitive events, the overall energy barriers would not vanish even at much higher shear stress. The dislocation motion additionally requires the successful activation of the shear-insensitive events with the assistance of thermal fluctuation. 
For partial~I, the activation of shear-sensitive and insensitive events takes place synchronously during the kink-pair nucleation. If one of these is unsuccessful due to insufficient external stress and/or temperature, then the dislocation becomes temporarily immobile, which is referred to as the self-pinning nature of synchro-Shockley dislocations in Laves phases \cite{kazantzis2007mechanical,kazantzis2008self}.
For partial~II, the thermal activation events could only dominate the transition process when the energy barrier of the athermal process is significantly reduced by external shear stress. In addition, unlike in partial~I, the shear-sensitive and insensitive events of partial~II motion are activated separately, thus, the transition does not require consecutive activation of events with different energy barriers.

In addition to the differences in energy barriers of thermally activated steps, the transition processes of the motion of partial~I and II show different stress responses (Figure \ref{fig2} and \ref{fig3}). A cross-over of partial~I and II glide-dominant regimes is thus expected as the temperature and/or stress level changes. For a high enough applied shear stress, partial~II is expected to have a higher mobility than partial~I at intermediate temperature. It is so because the non-sequential shuffling becomes the RLS of partial~II, which exhibits a lower activation energy than the thermally activated steps of partial~I. At low-stress levels but high temperatures, partial~I should be more mobile than partial~II. In fact, the thermal fluctuation can assist the motion of partial~I by overcoming the barriers of thermally activated steps, but not the motion of partial~II, where the shear straining as an athermal process brings the highest energy difference among all transition processes. At low temperatures when thermal fluctuations are not sufficient in activating synchro-Shockley dislocations, dislocation glide between triple and kagom\'e layers is expected to be the dominant dislocation event at high-stress levels.

The overall activation volume for the motion of synchro-Shockley dislocations consists of a shear-insensitive component, i.e., the volume for the short-range shuffling of small atoms in the triple-layer, and a shear-sensitive component obtained from the variation of activation energy versus applied stress.
For partial~I in this work, the activation volume of the shear-sensitive thermally activated event on the MEP is close to 2$b^3$. The activation volume for the shear-insensitive event is suspected to be similar. Therefore, an estimation of the activation volume of kink-pair nucleation with a few $b^3$ is reasonable and correlates well with the experimental estimations on C14 CaMg\textsubscript{2} ($\Omega < 15 b^3$, based on the experimental data obtained from micropillar compression tests \cite{freund2021plastic, zehnder2019plastic}) and other Laves phases \cite{kazantzis2007mechanical,kazantzis2008self,ohba1989high,saka1993plasticity}, where synchro-Shockley dislocations were believed to be the main carriers of plasticity.

Uniaxial applied tensile strain perpendicular to the slip plane can lower the energy barriers of synchro-Shockley dislocation motion, as the transition events involve out-of-plane atomic motion, see Figure \ref{fig2}, \ref{fig3} and S7. This is known as non-Schmid effect, namely the dependence of the Peierls barrier upon not only resolved shear stress but also normal stress, which is common in BCC metals for example. Interestingly, non-Schmid behavior has been recently reported in the micromechanical testing of M\textsubscript{n+1}AX\textsubscript{n} phases \cite{zhan2020non, zhan2021non} where the glide of zonal dislocations on the basal plane is believed to govern the plastic deformation \cite{plummer2022basal}. It is most probable that the non-Schmid behavior observed in MAX phases originates from alternative glide mechanisms similar or identical to those reported in this work. 

In summary, we show that thermal assistance is indispensable in the activation of the motion of one of the coupled synchro-Shockley partial dislocations. This finding provides an insight into the atomic origin of the brittle behavior in Laves phases upon deformation at ambient temperature due to lack of basal or $\{ 1 1 1 \}$ plasticity. 


\section*{Acknowledgments}
The authors acknowledge financial support by the Deutsche Forschungsgemeinschaft (DFG) through the projects A02, A05 and C02 of the SFB1394 Structural and Chemical Atomic Complexity – From Defect Phase Diagrams to Material Properties, project ID 409476157. This project has received funding from the European Research Council (ERC) under the European Union’s Horizon 2020 research and innovation programme (grant agreement No. 852096 FunBlocks). J.G. acknowledges funding from the French National Research Agency (ANR), Grant ANR-21-CE08-0001 (ATOUUM). Simulations were performed with computing resources granted by RWTH Aachen University under project (p0020267) and by the EXPLOR center of the Université de Lorraine and by the GENCI-TGCC (Grant 2020-A0080911390).


\bibliography{main}
\clearpage

\onecolumngrid
\begin{center}
\Large Supplementary Material: \\Thermally activated nature of synchro-Shockley dislocations in Laves phases\par
\end{center}

\setcounter{figure}{0}  
\renewcommand{\figurename}{Figure S}

\begin{figure}[hpt!]
\centering
\includegraphics[width=0.5\textwidth]{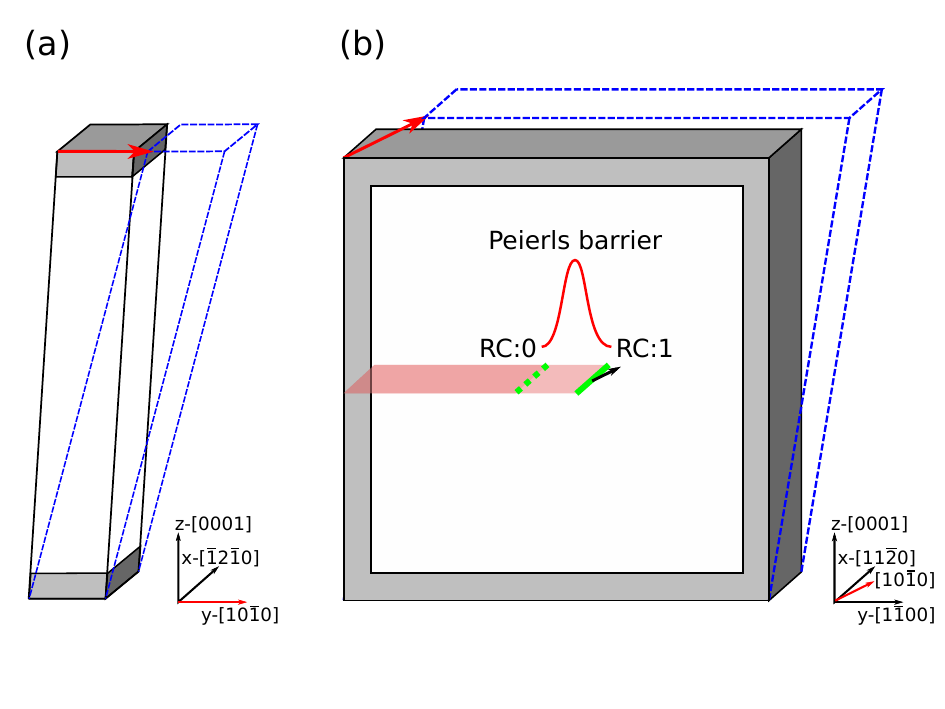}
\caption{Schematic illustration of the simulation setup. (a) Simulation setup for the strain-controlled nudged elastic band (NEB) calculation of minimum energy path (MEP) of synchro-shear slip. The dimensions of the setup in the periodic directions $x$ and $y$ are 6 and 10 \AA, respectively. For the non-periodic direction $z$, the dimension is 200 \AA.  (b) Slab setup for the NEB calculation of MEP of synchro-Shockley dislocation motion. 
The dimensions of the slab in non-periodic directions are around 300 \AA. 
Periodic boundary conditions are applied along the dislocation line direction ($x$) with a dimension of 61 \AA. The green line indicates the dislocation line and the black arrow indicates the Burgers vector. Please note that the shown setup corresponds to partial~I, for partial~II the stacking fault is on the right of the dislocation.
In (a) and (b), semi-fixed outer layers where atoms are frozen in non-periodic directions are marked in grey, and the thickness is more than 2 times the interatomic potential cutoff ($>$ 14 \AA). The red arrow indicates the transverse displacement and the blue dashed box indicates the sample with applied shear strain. The crystallographic orientation (x-$[1 1 \bar{2} 0]$, y-$[1 \bar{1} 0 0]$, and z-$[0001]$) was selected to investigate 30° synchro-Shockley dislocations, which have a dislocation line along x-$[1 1 \bar{2} 0]$ and glide along y-$[1 \bar{1} 0 0]$ or $[\bar{1} 1 0 0]$ on the basal plane (0001).}
\label{figS1}
\end{figure}

\begin{figure}[hpt!]
\centering
\includegraphics[width=0.5\textwidth]{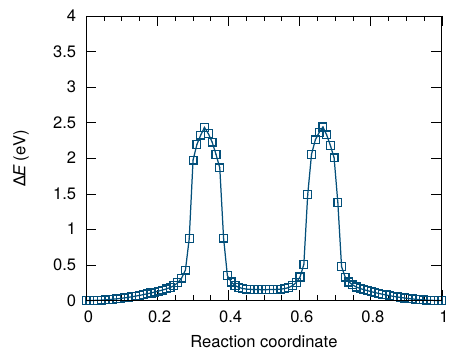}
\caption{Minimum energy path for synchro-shear full slip in C14 CaMg\textsubscript{2} at zero applied stress states.}
\label{figS2}
\end{figure}

\begin{figure*}[hptb!]
\centering
\includegraphics[width=0.9\textwidth]{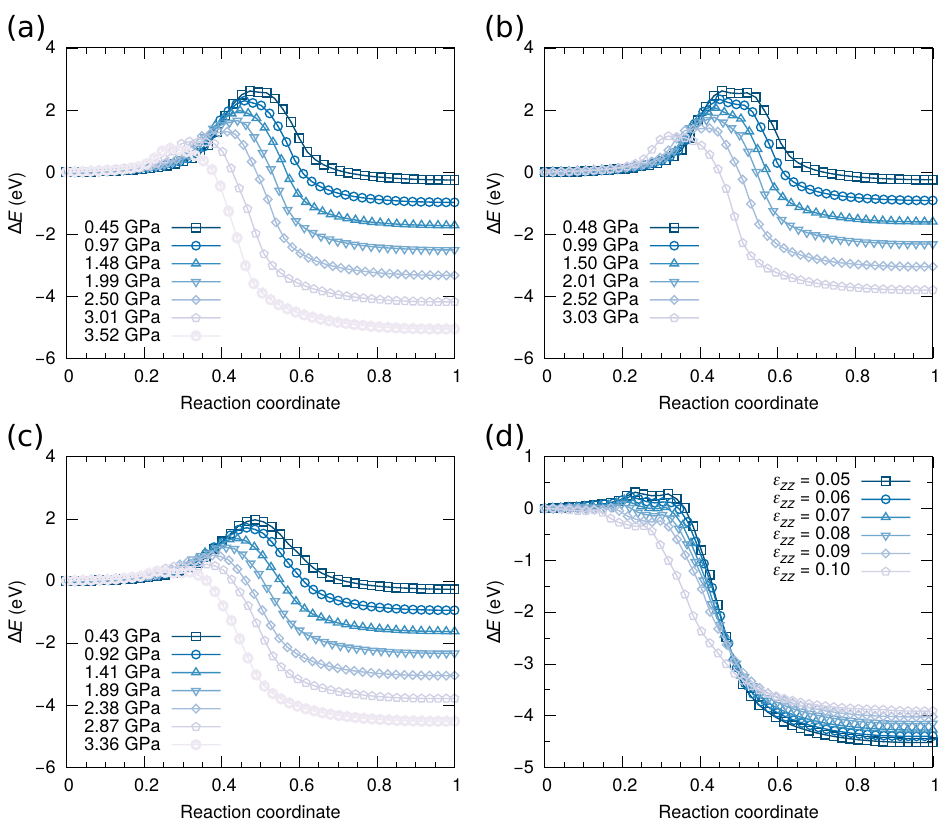}
\caption{Minimum energy paths for leading partial of synchro-shear slip at different shear and normal stress states. (a) Shear stress-dependent minimum energy paths for leading partial of synchro-shear slip under uniaxial tensile strain $\epsilon_{xx}$=0.05. (b) Shear stress-dependent minimum energy paths for leading partial of synchro-shear slip under uniaxial tensile strain $\epsilon_{yy}$=0.05. (c) Shear stress-dependent minimum energy paths for leading partial of synchro-shear slip under uniaxial tensile strain $\epsilon_{zz}$=0.05. (d) Normal strain-dependent minimum energy paths for leading partial of synchro-shear slip under shear strain $\epsilon_{yz}$=0.12.}
\label{figS3}
\end{figure*}
\FloatBarrier

\begin{figure*}[hptb!]
\centering
\includegraphics[width=0.9\textwidth]{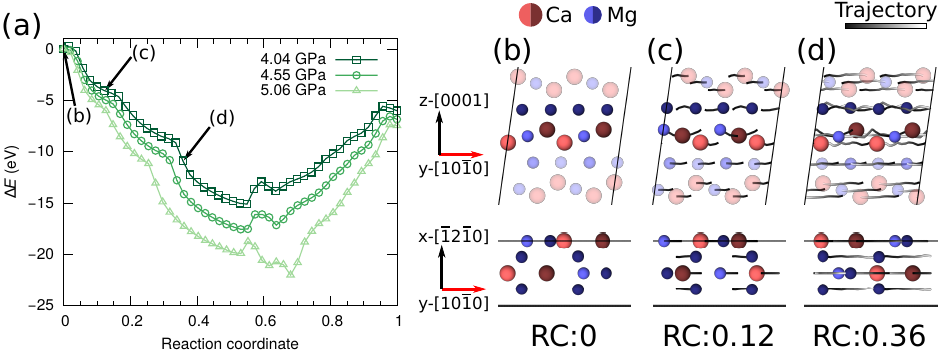}
\caption{Slip events in C14 CaMg\textsubscript{2} when $\tau$ $>$ 3.52 GPa. (a) Shear stress-dependent minimum energy paths for the slip events. (b-d) Snapshots of the slip events at $\tau =$ 4.04 GPa. Only atoms in the triple-layer and the adjacent kagomé layer (highlighted) are shown in the view along $[0001]$ (bottom). Large (red) and small (blue) spheres are Ca and Mg atoms, respectively. Dark and light atoms indicate atoms in different atomic layers. Grey-scale lines indicate the trajectories of atoms colored according to the reaction coordinates. The transverse displacement vectors (in red) and corresponding crystallographic orientations are indicated in the coordinate system.}
\label{figS4}
\end{figure*}

\begin{figure*}[hptb!]
\centering
\includegraphics[width=0.9\textwidth]{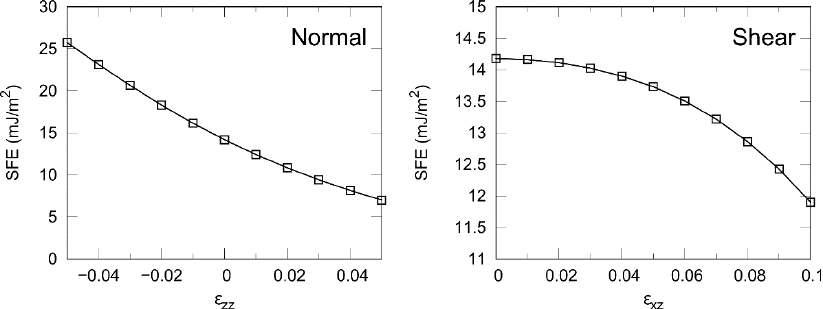}
\caption{Strain-dependent (normal or shear strain) stacking fault energy of the simulated C14 CaMg\textsubscript{2}. The stacking fault energy decreases with increasing tensile and shear strain. The stacking fault energies in Laves phases are expected to depend on temperature and applied stress. However, when comparing the energy contribution of the stacking fault extension with the energy required for kink-pair nucleation, this effect is negligible. E.g., in the kink pair nucleation process at 0\% applied strain, the energy required for the extension of stacking fault is only 1.5\% (0.034 eV, considering the width of the kink-pair is 7.3 Å, see Figure S8(b)) of the overall energy barrier (2.20 eV), which is very small.}
\label{figS5}
\end{figure*}

\begin{figure*}[hptb!]
\centering
\includegraphics[width=0.9\textwidth]{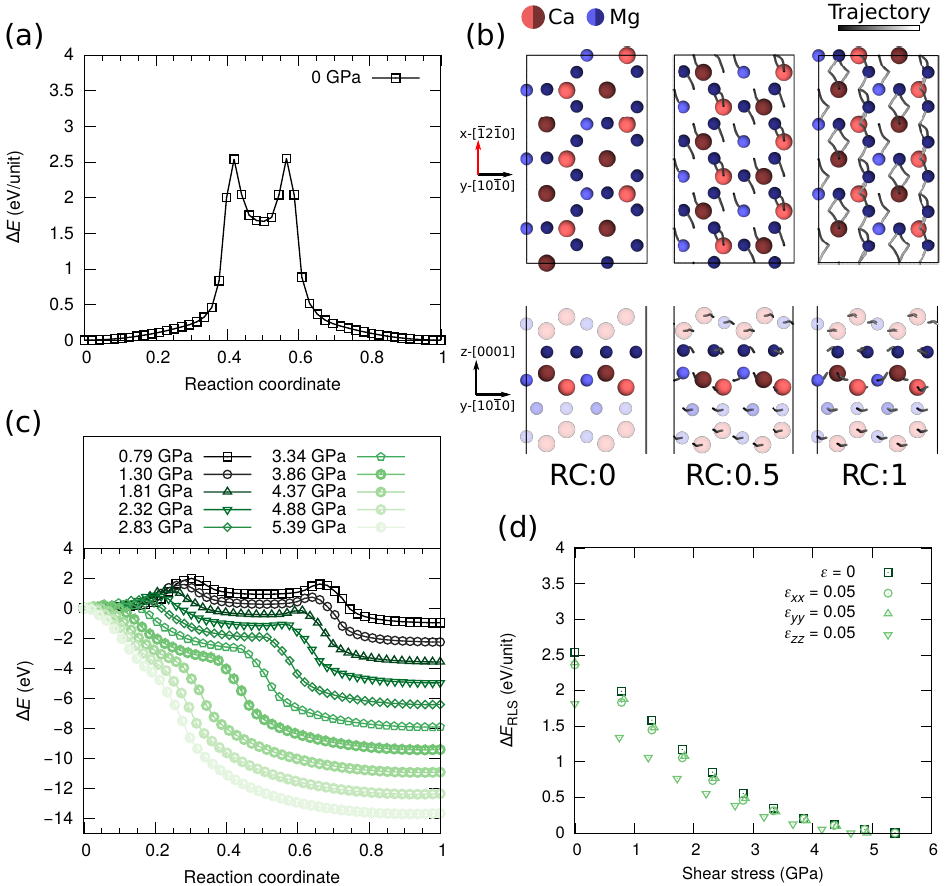}
\caption{Triple-kagomé full slip in C14 CaMg\textsubscript{2}. (a) Minimum energy path for triple-kagomé full slip in C14 CaMg\textsubscript{2} at zero applied stress states. (b) Snapshots of the triple-kagomé full slip mechanisms. Only atoms in the triple-layer and the adjacent kagomé layer (highlighted) are shown in the view along $[0001]$ (bottom). Large (red) and small (blue) spheres are Ca and Mg atoms, respectively. Dark and light atoms indicate atoms in different atomic layers. Grey-scale lines indicate the trajectories of atoms colored according to the reaction coordinates. The transverse displacement vectors (in red) and corresponding crystallographic orientations are indicated in the coordinate system. (c) Shear stress-dependent minimum energy paths for the triple-kagomé full slip. (d) Shear stress-dependent activation energy of rate-limiting step (RLS) of the triple-kagomé full slip.}
\label{figS6}
\end{figure*}

\begin{figure*}[hptb!]
\centering
\includegraphics[width=0.9\textwidth]{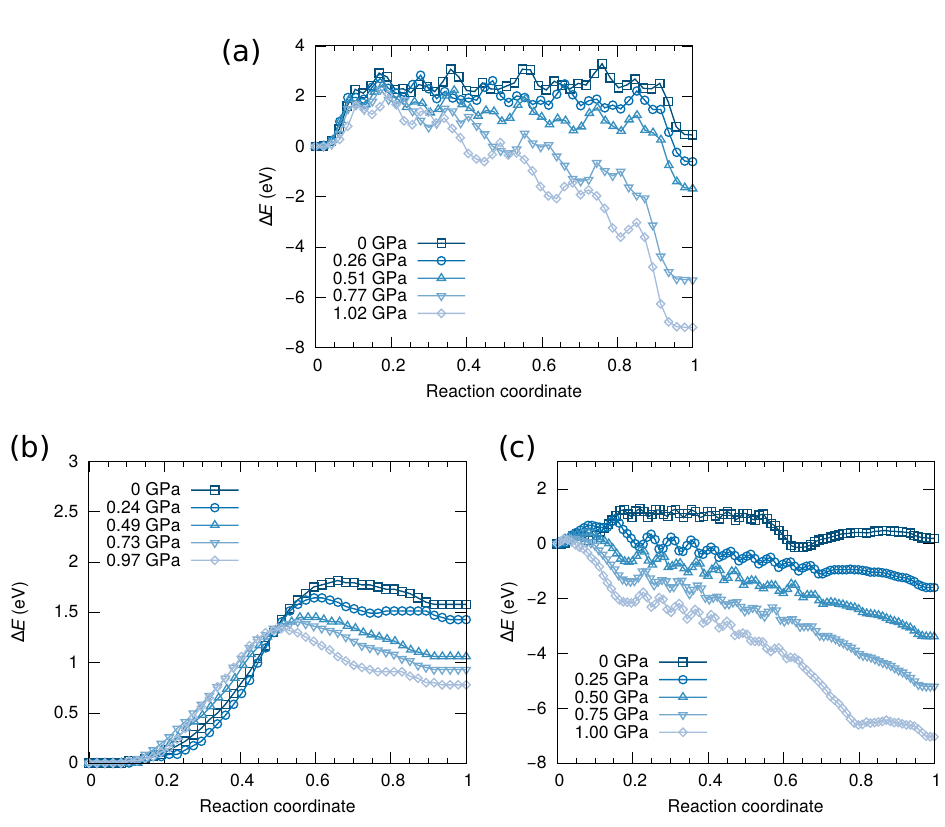}
\caption{Minimum energy paths for (a) motion of partial~I dislocation at different applied shear stress states and zero uniaxial tensile strain, (b) kink-pair nucleation of partial~I dislocation at different applied shear stress states and uniaxial tensile strain $\epsilon_{zz}$=0.05 and (c) motion of partial~II dislocation at different applied shear stress states and uniaxial tensile strain $\epsilon_{zz}$=0.02 in C14 CaMg\textsubscript{2}.}
\label{figS7}
\end{figure*}

\begin{figure*}[hptb!]
\centering
\includegraphics[width=0.6\textwidth]{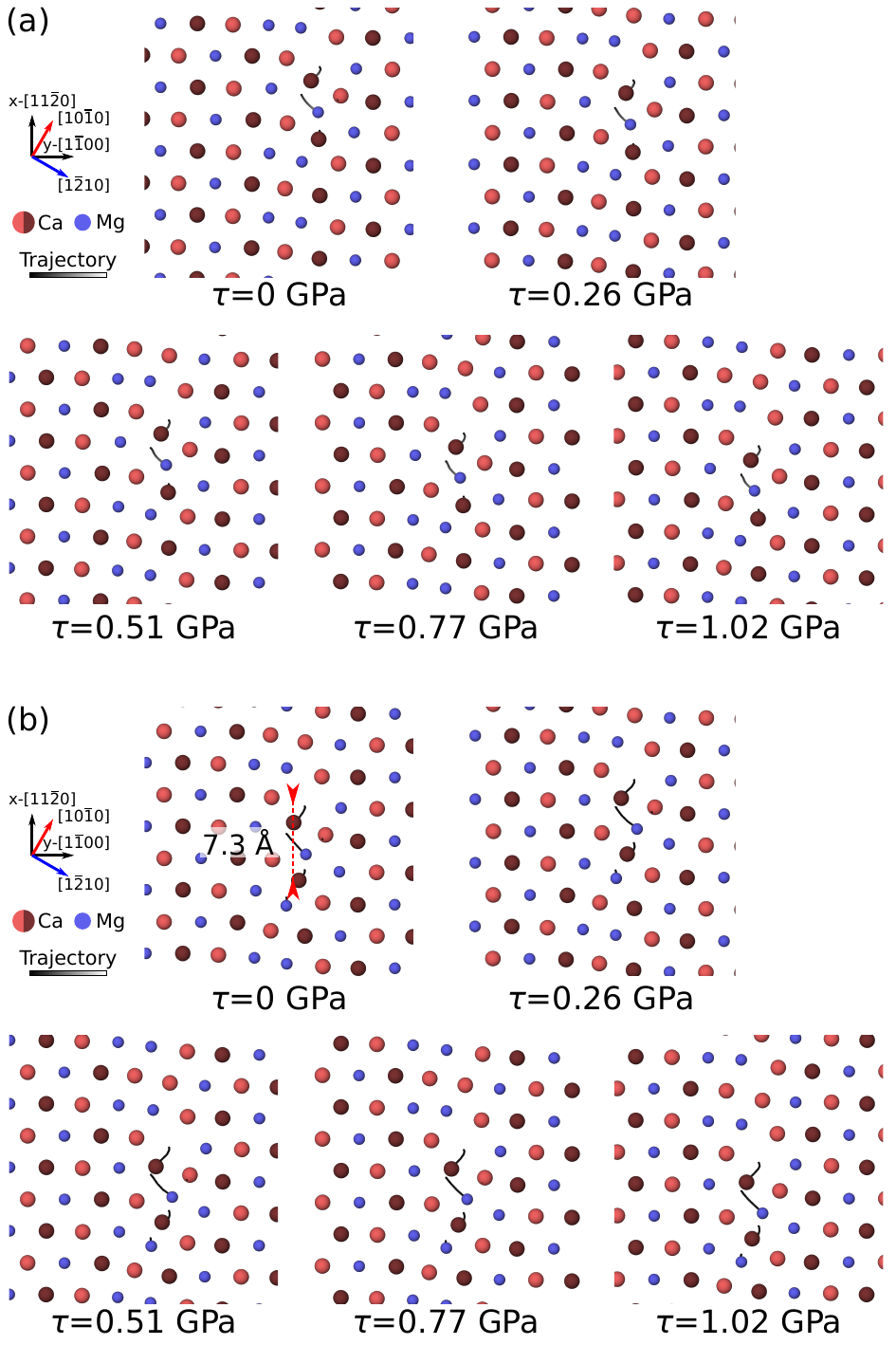}
\caption{Kink-pair nucleation of partial I dislocation in C14 CaMg\textsubscript{2} at different applied shear stress states ($\epsilon_{zz}$=0). (a) Snapshots at the energy barrier of the kink-pair nucleation process as illustrated in Figure 2(a). (b) Snapshots at the first local minima after the energy barrier of the kink-pair nucleation process as illustrated in Figure S7(a). The kink-pair distance is defined as the distance from the center of the vacancy to the interstial Ca atom. Only atoms in the triple-layer where the synchro-shear slip takes place are shown in the view along $[0001]$. Large (red) and small (blue) spheres are Ca and Mg atoms, respectively. Dark and light red atoms indicate Ca atoms in different atomic layers of the triple-layer. Grey-scale lines indicate the trajectories of atoms colored according to the reaction coordinates.}
\label{figS8}
\end{figure*}

\FloatBarrier
\clearpage

\end{document}